\documentclass{evn2004}
\usepackage{txfonts}
\begin{document}
   \title{The global properties of all variety of AGN}

   \author{M.J.M. March\~a}

   \institute{CAAUL, Observat\'orio Astron\'omico de Lisboa, Tapada da
             Ajuda, 1349-018 Lisboa, Portugal }

   \abstract{Active Galactic Nuclei (AGN) have been a challenging
   field of research for the past six decades. Nevertheless, many
   questions still remain unanswered today, regardless of the
   tremendous theoretical and technological advances. In this brief review
   I propose to take a step back from the usual discussion of AGN
   properties and draw attention to some topics that I believe are
   important to keep in mind as we strive forward in our pursuit of
   knowledge about these sources.

 }

   \maketitle
%

\section{Foreword}

The first class of Active Galactic Nuclei (AGN) was recognised over
six decades ago when Carl Seyfert selected and studied a group of
galaxies with unusual nuclear properties. Since then observational
windows beyond the visual boundaries have been opened into the
Universe and consequently a variety of phenomena have been discovered
and catalogued. Complementary to the technological development, great
steps forward have been made in both theory and interpretation
providing the concept that unification of ultimately all AGN types is
attainable. What I find interesting is that over the decades a
circular pattern seems to have developed: a technological advancement,
which is both possible and necessary by previous work, leads to the
discovery of a new phenomenon requiring a massive investigation via
surveying large numbers of sources, which in turn provokes a new wave
of interpretation and theoretical modelling, thus prompting the
building of new instruments. The circular pattern is renewed, and with
each renewal we diversify the variety of phenomena, but at the same time our
confidence in a unified picture increases.

What I propose to do in this review is to discuss some steps along the
circular pattern in the search of understanding AGN that I am more
familiar with, and which I believe are important to keep in mind as we
strive forward. In particular, I will dedicate some time to the more
general question of what is it that constitutes an AGN, their
properties, and the ever present problem of selection effects. In the
final Section I will mention the search for a parameter space that
allows the unification  of all types of AGN. I would like to point out,
however, that I do not pretend to make a complete review of the
subject (as this would be impossible given the time and scope of the
present meeting), and apologise for leaving out many undoubtedly
important, aspects of AGN studies.


\section{What is an AGN? }

What are the basic criteria that lead an object to be classified
as an AGN?  This is more than a simple academic question, and one that
assumes particular relevance as we develop new and more sensitive
instruments offering closer insights into the properties of the
so-called `normal' and `active galaxies'.

Clearly what is understood by an AGN is dependent on the ever evolving
wealth of multi-wavelength data. Nevertheless, the underlying
characteristic of these sources is their extra activity across the
electromagnetic spectrum. For example, the galaxies studied by Carl
Seyfert were peculiar in that they showed broader emission lines than
the other, more usual, spiral galaxies. (In fact, finding broad
emission lines in the spectrum of a given source is one of the most
conclusive ways of determining the presence of an AGN.) A decade or so
later after Seyfert published his work on the galaxies with unusual
spectra, the advent of radio astronomy  disclosed that some galaxies
were strong emitters in this region of the electromagnetic
spectrum. As we know, the discovery of the first quasar was a direct
consequence of the optical identification of radio sources, and  the
subject of Active Galactic Nuclei was forever established.

Many years after, and many technological and theoretical steps
forward, a paradigm for AGN has formed (see Figure 1). Its basic
assertion is that AGN drive their power from the accretion of matter
onto a supermassive ($M_{bh} \geq 10^{5} M_{\odot}$) black hole. The
signature of the disc is picked up across the spectrum in the optical,
UV and X-rays. Broad emission lines are produced by gas moving rapidly
in the potential of the black hole, while, further away from the black
hole, slower moving gas is responsible for narrower line
emission. Because some sources fail to show broad emission lines,
except in polarized light, an obscuring torus or warped disc has been
assumed although its morphology and properties is still subject of
much work. In a subgroup (typically considered to be less than 20\%)
of the AGN population, strong radio emission is associated with
relativistic jets and radio lobes that can extend over many
kiloparsecs.

Even though Figure 1 summarises the acquired knowledge over decades of
work, the fact is that much remains to be done and understood. Two
comments should accompany the AGN diagram  of Figure
1. Firstly, even if we have identified the different
ingredients that are related to the AGN phenomena, we have yet to
understand fully how the different combinations arise, or how they
relate to each other. Why do some sources show the radio strong
structures and others do not? Do the properties of the obscuring
material vary with luminosity, type of source, redshift?  These are
just examples of questions that have persisted for years, in spite of
being the topic of continued research. The second comment is concerned
with the lack of spherical symmetry in AGN. Such lack of symmetry
means that depending on the relative orientation between observer and
source, the effects of projection, relativistic beaming, and
obscuration, the same object can appear significantly
different. Distinguishing between intrinsic differences and those
induced by selection effects is not a simple task, but one which is
always relevant when analysing AGN samples. This will be discussed in
the next Section.

%
   \begin{figure}
   \centering
\vspace{200pt}
   \includegraphics{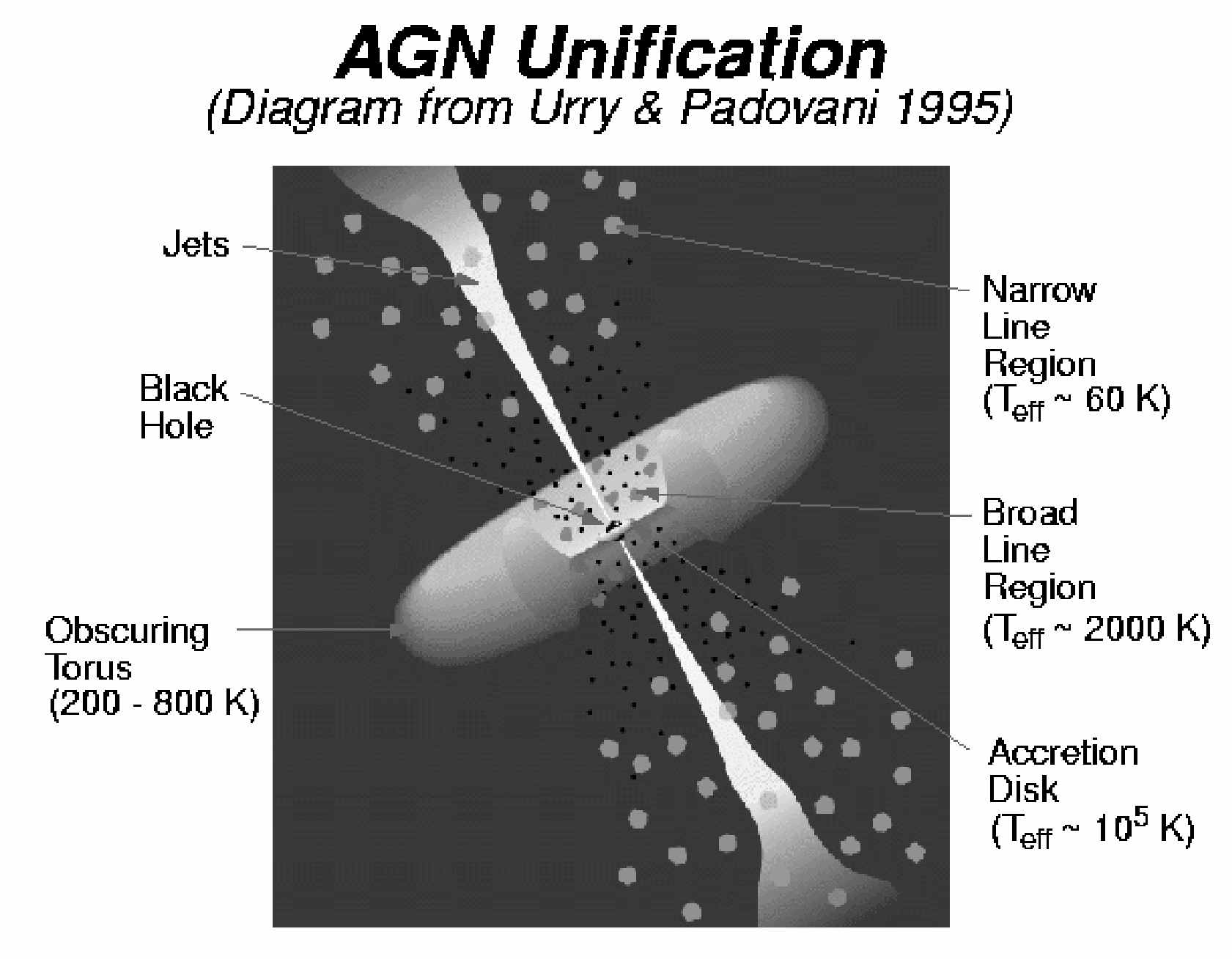}
      \caption{The AGN diagram is taken from Urry \& Padovani,
      (\cite{urry95}), and annotated by M. Voit. It shows the
      ingredients of all types of AGN: the central black hole and
      accretion disk, the broad line region and the obscuring torus
      that hides it from some directions of observation, the narrow
      line region further out from the black hole, and finally the
      radio jets and extended emission.
         \label{fig.agnmodel}
         }
   \end{figure}
%


\section{The properties of AGN }

Surveys and catalogues are inevitable steps in any scientific area
where quantitative results are pursued. The topic of AGN is no
exception. We want to know how many there are, are they all the same,
how are they distributed, how do the properties vary with cosmic time,
or with luminosity, or any other property?

In order to answer these questions it is first necessary to create some order
in the multitude of observed properties. Since optical and
radio surveys of AGN were the first to be carried out, a great deal of the
classification process is linked to the properties of sources in these two
regions of the electromagnetic spectrum. For instance, the
classification of Quasar was given to those sources that appeared
point like on the optical charts but whose spectra showed broad
emission lines, whereas the term Radio Galaxy was assigned to those
sources with strong radio emission but which appeared fuzzy in optical
catalogues. Many other classes and sub-classes of AGN are now in use
depending on the combination of properties such as detectability,
polarization, spectra and variability. However, in 1995 Urry \&
Padovani proposed a simplification of the AGN diversity by
suggesting three classes of sources according to the optical and UV
spectra: 

{\bf Type 1 AGN:} those with bright continua and broad emission lines
in their spectra. These include Seyfert type~1, Quasars and Broad Line
Radio Galaxies (BLRGs). According to the diagram of Figure 1, these
sources correspond to those that the observer sees at angles that
allow direct observation of the broad line region. 

{\bf Type 2 AGN:} those with weak continua and only narrow emission
lines in their spectra. These broadly include Seyfert type~2, Narrow
Line Radio Galaxies (NLRGs). In the framework of the AGN diagram of
Figure~1, these sources correspond to those objects which the observer
sees along the direction of the obscuring 'torus'.  

{\bf Type 0 AGN:} those sources which lack strong emission or
absorption features in the their spectra (BL Lacs), or those quasars
which are highly polarized (HPQ for Highly Polarized Quasar), or
extremely variable in the optical (the OVV for Optically Violently
Variable). According to the schematic model for an AGN shown in
Figure~1, these types of sources are thought to be those where the
direction of the radio jet is well aligned with the line of sight. 

Even though classification is an important stage in the pursuit of
knowledge, the real scientific quest lies beyond that. In particular, the
goal is is to try and look for relationships between classes and
distinguish between those properties that are intrinsic, and those
that are induced by the observing criteria and/or sensitivity of the
instruments.  In this Section I would like to draw attention to two
outstanding issues that can be used to illustrate this point.

   \begin{figure}
   \centering
   \vspace{250pt}
   \includegraphics{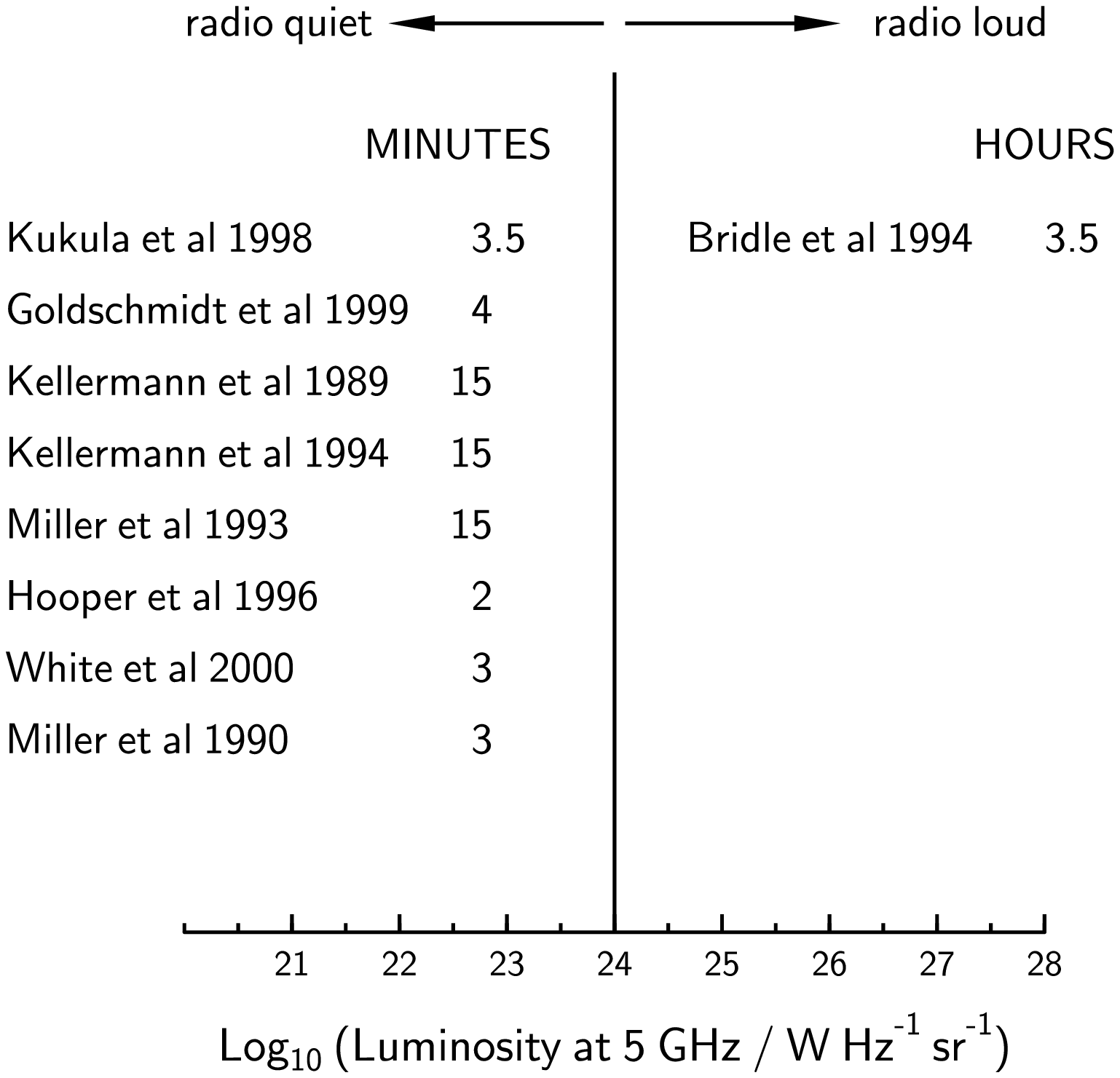}
      \caption{The figure is taken from Blundell (\cite{blundell03b}
      and it summarises the average integration time for observations
      of RQ and RL sources).
         \label{fig.blundell}
         }
   \end{figure}
%

{\bf Case 1: The radio-loud/radio-quiet bimodality.} The fact is that
there seems to be a split in the the Spectral Energy Distribution
(SED) of AGN at radio-frequencies, with the radio-quiet (RQ) AGN
lacking the relativistic jet emission and lobe structures seen in
their radio-loud (RL) counterparts. In order to discuss the subject
more objectively, empirical criteria were independently set to
separate the two sub-groups. Specifically, RQ AGN have been separated
by requiring that either their radio luminosity at 5~GHz $L_{5} <
10^{24}$~W~Hz$^{-1}$~sr$^{-1}$ (Miller et al., \cite{miller90}), or
that the ratio of radio to optical flux ($r = S_{5GHz}/S_{4400\AA} <
10$, Kellermann et al. 1994). Evidence for and against the bimodality
using a variety of surveys has been found by a number of authors
(White et al. 2000, Ivezic et al. 2002; Cirasuolo, et al 2003, just to
mention a few more recent works). The issue is obviously of extreme
importance for the understanding of the AGN phenomena. If there is
truly a bimodality rather than a continuity in the radio properties of
AGN, then we need to find out what is (are) the parameter(s)
responsible for setting on/off the radio emission mechanism. At this
point it may help to take a step back and ask the question of whether
the bimodality claimed by some works could be the result of
incompleteness or sensitivity related problems, rather than intrinsic
differences in the population. This exercise is of particular
relevance when recent claims that superluminal motion have been
discovered in a couple of sources classified as RQ (Brunthaler et
al. 2000, Blundell, et al., 2003a) according to the empirical criteria
mentioned before. A closer look at these criteria reveals some
potential problems with such sharp cutoffs. In the case of the ratio
$r$, the danger is that the same cutoff value of 10, when applied to
different samples, may introduce different biases. In general, the
optical flux contains the contributions from both the host galaxy, and
the AGN nucleus. However, the amounts of galaxy and nuclear light
contributing to the flux measured through a fixed size aperture varies
significantly with redshift. The consequence of this is that the
optical flux measured for a sample of quasars will be mostly nuclear
flux, while for a sample of close by objects the optical flux measured
will include a significant contribution of the host galaxy
light. Hence, the quantity $r$ will actually measure different things
in the two samples. This has been noted by Ho \& Peng (\cite{ho01}) who have pointed out
that a large fraction of Seyfert galaxies, usually considered to be
RQ, would actually become RL objects according to the $r$ criterion,
if only their nuclear contribution was considered for the optical
flux.

Looking now to the other parameter used to separate RL from RQ AGN, I
draw attention to Figure 2 of Blundell (\cite{blundell03b})  which is reproduced here
(Figure~\ref{fig.blundell}) where a simple listing of mean length of
time-on-source for observations of RQ and RL sources is put side by
side. It is striking to remark that the observations referring to the
RQ sources range in the units of minutes, while those referring to the
RL sources are in units of hours. Moreover, the author points out that
in addition to the significantly shorter integration times for the RQ
sources, these observations were usually made using the most extended
VLA configurations which are less sensitive to low surface brightness,
extended emission. As a conclusion, I would like to point out that it
is remarkable that such a long standing issue in AGN such as the RQ/RL
dichotomy/continuity is still plagued, not so much by the lack of good
quality data, but rather by selection effects.

\begin{figure}
   \centering
   \vspace{280pt}
   \includegraphics{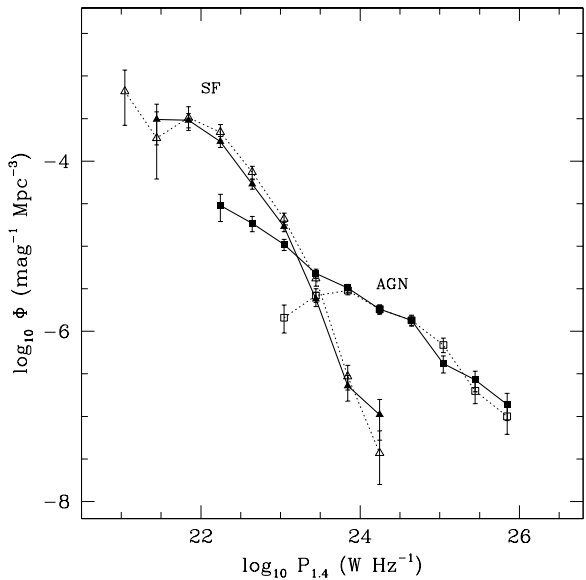}
      \caption{The RLF from Sadler et al. (\cite{sadler02}). The
      triangles refer to star forming galaxies and squares to AGN; the
      dashed lines refer to the estimate of Machalski \& Godlowski
      (\cite{machalski00}).).
         \label{fig.sadler}
         }
   \end{figure}

{\bf Case 2: The Radio Luminosity Function (RLF).}  One of the
fundamental questions about AGN in general, but also about the
different classes, is how many there are, how do they distribute
themselves, and how did they come into being. Faced with the
impossibility of following the life cycle of such a source, the best
we can achieve is a statistical description of the data. A usual way
of doing this is to determine the luminosity function (LF) that
basically describes how the comoving number density of sources varies
with luminosity. Traditionally, radio surveys have been preferred
since radio waves are unaffected by dust obscuration, hence the
determination of RLF for AGN. Two of the most recent determinations of
the RLF at 1.4~GHz considering large number of sources ($>$500) are
shown in Figure~\ref{fig.sadler} (taken from Sadler et al. 2002). The
triangles represent star forming galaxies (SF), and the squares the
AGN. The full line represents the data from Sadler et al., and the
dashed line that of Machalski \& Godlowski (2000). 

A couple of noteworthy points should be made about these results. The
first is that the SF galaxies and AGN contribute significantly to the
local RLF for radio powers below 10$^{24}$~W~Hz$^{-1}$. The second
point is that the agreement between the two works represented in
Figure~\ref{fig.sadler} over much of the range of luminosities covered, 
breaks down when the samples are split into SF and AGN. In the
estimate of Machalski \& Godlowski 2000 (dashed line) the contribution
of SF galaxies decreases below radio powers of roughly 10$^{24}$~W/Hz,
whereas in the estimate by Sadler et al.  (full line) this
contribution continues to rise all the way radio powers of $\sim
10^{22}$~W/Hz. Can this discrepancy be the consequence of selection
effects? Sadler et al. (2002) actually discusses briefly the possible
reasons for this divergence in the contribution of SF galaxies at low
radio powers. Their conclusion is that both selection differences and
misclassification of low power AGN as SF galaxies can be responsible
for incompleteness in the Machalski \& Godlowski data. Such conclusion
brings to the fore, once again, the importance of identification and
classification of sources. 

Basic difficulties such as being able to
separate and recognise AGN activity will be particularly relevant in
forthcoming surveys at the $\mu$-Jy level where extrapolations of the
local RLF suggest that a mixture of AGN and SF galaxies will be found
at all redshifts (see Sadler et al. 2002 for references).

%


\section{Searching for the parameter space where unification is possible}

The idea that the diversity of AGN can be understood via a reduced
number of parameters has been around for some time now. The first
suggestions of a unifying scheme had obscuration and orientation as
fundamental parameters (Rowan-Robison 1976, Lawrence \& Elvis,
1982). The discovery of superluminal motion in flat radio spectrum
sources, and its consequent interpretation in terms of bulk
relativistic motion, implied that the appearance of a radio-loud
object could change dramatically depending on the orientation. Hence,
obscuration, orientation, and Doppler boosting had to be taken into
account in order to unify all types of AGN.

In general, unification is discussed in two parallel schemes: one for
the radio-quiet, where Seyfert~1 and Seyfert~2 galaxies are
interpreted as intrinsically the same but where the latter are seen
along a direction that hides the broad line region behind a `torus'
(Antonucci \& Miller, 1985). On the other hand, the scheme for
radio-loud sources, needs to complement orientation and obscuration
with the effects of Doppler boosting in order to unify the
lobe-dominated and core-dominated sources (Orr \& Browne, 1982;
Barthel, 1989; Urry \& Padovani, 1995).

   \begin{figure}
   \centering
   \vspace{280pt}
   \includegraphics{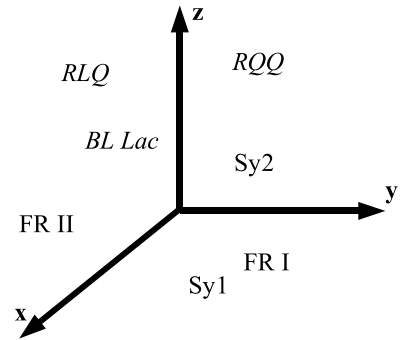}
      \caption{ Schematic representation of the parameter space to
      unify the AGN diversity. The challenge is to establish $x, y, z$
      as the parameters that control the life cycle of and AGN.).
         \label{fig.us}
         }
   \end{figure}
%

Despite the relative success of the unification schemes for the AGN
diversity, there is a growing feeling that a broader unification is
necessary if we are to understand the phenomena in its entirety. The
realisation that all galactic nuclei contain a black hole whose mass
is closely related to macroscopic quantities such as the luminosity
and the stellar velocity dispersion (Magorrian, et al. 1998; Gebhardt,
et al. 2000; merritt \& Ferrarese, 2001) gave a new impetus to the
concept of unification schemes. In fact, schemes are no longer just
sought for understanding AGN diversity `per se'. The goal is now to
obtain a `grand unifying scheme' that can offer a 'global' connection
between AGN and galaxy formation and evolution, since AGN activity
seems inevitably related to a phase (or phases), in the life cycle of
a galaxy. What we seek now is to identify the axis {\bf $x,y,z$} of a
parameter space (see Figure~\ref{fig.us}) where the AGN diversity is
understood in terms of an evolutionary process that involves the
characteristics of the black hole at the centre, and its immediate and
more ample environment. 

The present knowledge about AGN and black hole physics gives hints as
to what parameters will play a role: mass accretion rate seems to be
extremely important for the emission properties of the central source
(Meier, 2002), and it may be responsible for triggering on/off the AGN
activity or even a sequence in the activity (Urry, 2003; Cavaliere \&
d'Elia, 2002; Cao, 2003). On the other hand, black hole spin has long
been deemed relevant for the jet production (Blandford \& Znajek,
1977; Meier 2002 for references), and though much work seems to be
required especially in what concerns the magnetic field, it is likely
that this characteristic will have a role to play in the grand
unification scheme of AGN. Finally, environment and interaction
between galaxies is bound to influence significantly the observed
properties of AGN, as they may dictate when and how this 'active'
phase is triggered. As a concluding remark, I would argue that finding
the axis of the parameter space that trace the AGN/galaxy evolution,
and diversity, is one of the biggest and challenging endeavours in the
study of Active Galactic nuclei.

\begin{acknowledgements}
The author acknowledges the financial support of the Funda\c{c}\~ao
para a Ci\^encia e Tecnologia.

\end{acknowledgements}

\cleardoublepage

\end{document}